\documentclass[a4paper]{jpconf}
\usepackage{graphicx}
\begin{document}

\newcommand{\beq} {\begin{equation}}
\newcommand{\enq} {\end{equation}}
\newcommand{\ber} {\begin {eqnarray}}
\newcommand{\enr} {\end {eqnarray}}
\newcommand{\eq} {equation}
\newcommand{\eqs} {equations }
\newcommand{\mn}  {{\mu \nu}}
\newcommand{\sn}  {{\sigma \nu}}
\newcommand{\rhm}  {{\rho \mu}}
\newcommand{\sr}  {{\sigma \rho}}
\newcommand{\bh}  {{\bar h}}
\newcommand {\er}[1] {equation (\ref{#1}) }
\newcommand {\ern}[1] {equation (\ref{#1})}
\newcommand{\mbf} {{ }}
\newcommand {\Er}[1] {Equation (\ref{#1}) }

\title{A New Diffeomorphism Symmetry Group of Non-Barotropic Magnetohydrodynamics}

\author{Asher Yahalom}

\address{Ariel University, Ariel 40700, Israel}

\ead{asya@ariel.ac.il}

\begin{abstract}
The theorem of Noether dictates that for every continuous symmetry group of an Action the system must possess a conservation law.
In this paper we discuss some subgroups of Arnold's labelling symmetry diffeomorphism related to non-barotropic magnetohydrodynamics (MHD) and the conservations laws associated with them. Those include but are not limited to the metage translation group and the associated topological conservations law of non-barotropic cross helicity.
\end{abstract}

\section{Introduction}
\label{sec:1}

The theorem of Noether dictates that for every continuous symmetry group of an Action the system must possess a conservation law.
For example time translation symmetry results in the conservation of energy, while spatial translation symmetry results in the conservation of linear momentum and rotation symmetry in the conservation of angular momentum to list some well known examples. But sometimes the conservation law is discovered without reference to the Noether theorem by using the equations of the system. In that case one is tempted to inquire what is
the hidden symmetry associated with this conservation law and what is the simplest way to represent it.

The concept of metage as a label for fluid elements along a vortex line in ideal fluids was first introduced by
Lynden-Bell \& Katz \cite{LynanKatz}. A translation group of this label was found to be connected to the conservation
of Moffat's \cite{Moffatt} helicity by Yahalom \cite{Yahalomhel}. The concept of metage was later generalized by Yahalom \& Lynden-Bell \cite{YaLy} for barotropic MHD, but now as a label for fluid elements along magnetic field lines which are comoving with the flow in the case of ideal MHD. Yahalom \& Lynden-Bell \cite{YaLy} has also shown that the translation group of the magnetic metage is connected to Woltjer \cite{Woltjer1,Woltjer2} conservation of cross helicity for barotropic MHD. Recently the concept of metage was generalized also for non barotropic MHD in which magnetic field lines lie on entropy surfaces \cite{simpvarYah}. This was later generalized by dropping the entropy condition on magnetic field lines \cite{metra}.

Cross Helicity was first described by Woltjer \cite{Woltjer1,Woltjer2} and is give by:
\begin{equation} \label{GrindEQ__22_}
H_{C} \equiv \int  \vec{B}\cdot \vec{v}d^{3} x,
\end{equation}
in which $\vec{B}$ is the magnetic field, $\vec{v}$ is the velocity field and the integral is taken
over the entire flow domain. $H_{C}$ is conserved for barotropic or incompressible MHD and
is given a topological interpretation in terms of the knottiness of magnetic and flow field lines.
A generalization of barotropic fluid dynamics conserved quantities including helicity to non barotropic flows
including topological constants of motion is given by Mobbs \cite{Mobbs}. However, Mobbs did not
discuss the MHD case.

Both conservation laws for the helicity in the fluid dynamics case and the barotropic MHD case were
shown to originate from a relabelling symmetry through the Noether theorem \cite{Yahalomhel,Padhye1,Padhye2,YaLy}.
Webb et al. \cite{Webb2} have generalized the idea of relabelling symmetry to non-barotropic MHD and derived their generalized cross helicity conservation law by using Noether's theorem but without using the simple representation which is connected with the metage variable. The conservation law deduction involves a divergence symmetry of the action. These conservation laws were written as Eulerian conservation laws of the form $D_t+\vec \nabla \cdot \vec F = 0$ where D is the conserved density and F is the conserved flux. Webb et al. \cite{Webb4} discuss the cross helicity conservation law for non-barotropic MHD in a multi-symplectic formulation of MHD. Webb et al. \cite{Webb1,Webb2} emphasize that the generalized cross helicity conservation law, in MHD and the generalized helicity conservation law in non-barotropic fluids are non-local in the sense that they depend on the auxiliary nonlocal variable $\sigma$, which depends on the Lagrangian time integral of the temperature $T(x, t)$. Notice that a potential vorticity conservation equation for non-barotropic MHD is derived by Webb, G. M. and Mace, R.L. \cite{Webb5} by using Noether's second theorem.

It should be mentioned that Mobbs \cite{Mobbs} derived a helicity conservation law for ideal, non-barotropic fluid dynamics, which is of the same form as the cross helicity conservation law for non-barotropic MHD, except that the magnetic field induction is replaced by the generalized fluid helicity  $ \vec \Omega = \vec \nabla \times (\vec v - \sigma \vec \nabla s) $. Webb et al. \cite{Webb1,Webb2} also derive the Eulerian, differential form of Mobbs \cite{Mobbs} conservation law (although they did not reference Mobbs \cite{Mobbs}). Webb and Anco \cite{Webb3} show how Mobbs conservation law arises in multi-symplectic, Lagrangian fluid mechanics.

Variational principles for magnetohydrodynamics were introduced by
previous authors both in Lagrangian and Eulerian form. Sturrock
\cite{Sturrock} has discussed in his book a Lagrangian variational
formalism for magnetohydrodynamics.
Vladimirov and Moffatt \cite{VMoffatt} in a series of papers have discussed an Eulerian
variational principle for incompressible magnetohydrodynamics.
However, their variational principle contained three more
functions in addition to the seven variables which appear in the
standard equations of incompressible magnetohydrodynamics which are the magnetic
field $\vec B$ the velocity field $\vec v$ and the pressure $P$.
Kats \cite{Kats} has generalized Moffatt's work for compressible
non barotropic flows but without reducing the number of functions
and the computational load.  Sakurai \cite{12} has introduced a two function Eulerian variational
principle for force-free magnetohydrodynamics and used it as a
basis of a numerical scheme, his method is discussed in a book by
Sturrock \cite{Sturrock}. Yahalom \& Lynden-Bell \cite{YaLy} combined the Lagrangian of
Sturrock \cite{Sturrock} with the Lagrangian of Sakurai
\cite{12} to obtain an {\bf Eulerian} Lagrangian principle for barotropic magnetohydrodynamics
which will depend on only six functions. The variational
derivative of this Lagrangian produced all the equations
needed to describe barotropic magnetohydrodynamics without any
additional constraints. The equations obtained resembled the
equations of Frenkel, Levich \& Stilman \cite{FLS} (see also \cite{Zakharov}).
Yahalom \cite{Yah} have shown that for the barotropic case four functions will
suffice. Moreover, it was shown that the cuts of some of those functions \cite{Yah2}
are topological local conserved quantities.

Previous work was concerned only with
barotropic magnetohydrodynamics. Variational principles of non
barotropic magnetohydrodynamics can be found in the work of
Bekenstein \& Oron \cite{Bekenstien} in terms of 15 functions and
V.A. Kats \cite{Kats} in terms of 20 functions.
Morrison \cite{Morrison} has suggested a Hamiltonian approach but this also depends on 8 canonical variables (see table 2 \cite{Morrison}).
The variational principle introduced  in \cite{Yahalom1,Yahalom2} show that only five functions will suffice to
describe non barotropic MHD in the case that we enforce a Sakurai \cite{12} representation for the magnetic field.

The plan of this paper is as follows: First  we introduce the basic quantities and equations of non-barotropic MHD. Then we describe the concept of magnetic metage for non-barotropic MHD. This is followed by a description of a Lagrangian variational principle for non-barotropic MHD. Finally we discuss some subgroups of Arnold's labelling symmetry diffeomorphism related to non-barotropic MHD and the conservations laws associated with them.

\section{Basic Equations}

Consider the equations of non-barotropic MHD \cite{Sturrock,Yahalom1}:
\beq
\frac{\partial{\vec B}}{\partial t} = \vec \nabla \times (\vec v \times \vec B),
\label{Beq}
\enq
\beq
\vec \nabla \cdot \vec B =0,
\label{Bcon}
\enq
\beq
\frac{\partial{\rho}}{\partial t} + \vec \nabla \cdot (\rho \vec v ) = 0,
\label{masscon}
\enq
\beq
\rho \frac{d \vec v}{d t}=
\rho (\frac{\partial \vec v}{\partial t}+(\vec v \cdot \vec \nabla)\vec v)  = -\vec \nabla p (\rho,s) +
\frac{(\vec \nabla \times \vec B) \times \vec B}{4 \pi},
\label{Euler}
\enq
\beq
 \frac{d s}{d t}=\frac{\partial s}{\partial t}+(\vec v \cdot \vec \nabla) s = 0.
\label{Ent}
\enq
In the above the following notations are utilized: $\frac{\partial}{\partial t}$ is the temporal derivative,
$\frac{d}{d t}$ is the temporal material derivative and $\vec \nabla$ has its
standard meaning in vector calculus. $\rho$ is the fluid density and $s$ is the specific entropy. Finally $p (\rho,s)$ is the pressure which
depends on the density and entropy (the non-barotropic case).  \Er{Beq}describes the
fact that the magnetic field lines are moving with the fluid elements ("frozen" magnetic field lines),
 \ern{Bcon} describes the fact that the magnetic field is solenoidal, \ern{masscon} describes the conservation of mass and \ern{Euler}
is the Euler equation for a fluid in which both pressure
and Lorentz magnetic forces apply. \Er{Ent} describes the fact that heat is not created (zero viscosity, zero resistivity)
 in ideal non-barotropic MHD and is not conducted, thus only convection occurs.
The number of independent variables for which one needs to solve is eight
($\vec v,\vec B,\rho,s$) and the number of \eqs (\ref{Beq},\ref{masscon},\ref{Euler},\ref{Ent}) is also eight.
Notice that \ern{Bcon} is a condition on the initial $\vec B$ field and is satisfied automatically for
any other time due to \ern{Beq}.

In non-barotropic MHD one can calculate the temporal derivative of the cross helicity (\ref{GrindEQ__22_}) using the
above equations and obtain:
\begin{equation} \label{GrindEQ__22c_}
\frac{dH_{C}}{dt}= \int T  \vec{\nabla} s \cdot \vec{B} d^{3} x,
\end{equation}
in which $T$ is the temperature. Hence, generally speaking cross helicity is not conserved.

\section{Load and Metage}
\label{inverse}

The following section follows closely a similar section in \cite{YaLy}. Consider a thin tube surrounding a magnetic field line,
the magnetic flux contained within the tube is:
\beq
\Delta \Phi =
\int \vec B \cdot d \vec S
\label{flux}
\enq
and the mass
contained with the tube is:
\beq
\Delta M = \int \rho d\vec l \cdot d \vec S,
\label{Mass}
\enq
in which $dl$ is a length element
along the tube. Since the magnetic field lines move with the flow
by virtue of \ern{Beq} and \ern{masscon} both the quantities $\Delta \Phi$ and
$\Delta M$ are conserved and since the tube is thin we may define
the conserved magnetic load:
\beq
\lambda = \frac{\Delta M}{\Delta
\Phi} = \oint \frac{\rho}{B}dl,
\label{Load}
\enq
in which the
above integral is performed along the field line. Obviously the
parts of the line which go out of the flow to regions in which
$\rho=0$ have a null contribution to the integral.
Notice that $\lambda$ is a {\bf single valued} function that can be measured in principle.
Since $\lambda$
is conserved it satisfies the equation:
\beq
 \frac{d \lambda }{d t} = 0.
\label{Loadcon}
\enq
By construction surfaces of constant magnetic load move with the flow and contain
magnetic field lines. Hence the gradient to such surfaces must be orthogonal to
the field line:
\beq
\vec \nabla \lambda \cdot \vec B = 0.
\label{Loadortho}
\enq
Now consider an arbitrary comoving point on the magnetic field line and denote it by $i$,
and consider an additional comoving point on the magnetic field line and denote it by $r$.
The integral:
\beq
\mu(r)  = \int_i^r \frac{\rho}{B}dl + \mu(i),
\label{metage}
\enq
is also a conserved quantity which we may denote following Lynden-Bell \& Katz \cite{LynanKatz}
as the magnetic metage. $\mu(i)$ is an arbitrary number which can be chosen differently for each
magnetic line. By construction:
\beq
 \frac{d \mu }{d t} = 0.
\label{metagecon}
\enq
Also it is easy to see that by differentiating along the magnetic field line we obtain:
\beq
 \vec \nabla \mu \cdot \vec B = \rho.
\label{metageeq}
\enq
Notice that $\mu$ will be generally a {\bf non single valued} function, we will show later in this paper
that symmetry to translations in $\mu$; will generate through the Noether theorem the conservation of the
magnetic cross helicity.

At this point we have two comoving coordinates of flow, namely $\lambda,\mu$ obviously in a
three dimensional flow we also have a third coordinate. However, before defining the third coordinate
we will find it useful to work not directly with $\lambda$ but with a function of $\lambda$.
Now consider the magnetic flux within a surface of constant load $\Phi(\lambda)$. The magnetic flux is a conserved quantity
and depends only on the load $\lambda$ of the surrounding surface. Now we define the quantity:
\beq
 \chi = \frac{\Phi(\lambda)}{2\pi}.
\label{chidef}
\enq
Obviously $\chi$ satisfies the equations:
\beq
\frac{d \chi}{d t} = 0, \qquad \vec B \cdot \vec \nabla \chi = 0.
\label{chieq}
\enq
Let us now define an additional comoving coordinate $\eta^{*}$
since $\vec \nabla \mu$ is not orthogonal to the $\vec B$ lines we can choose $\vec \nabla \eta^{*}$ to be
orthogonal to the $\vec B$ lines and not be in the direction of the $\vec \nabla \chi$ lines,
that is we choose $\eta^{*}$
not to depend only on $\chi$. Since both $\vec \nabla \eta^{*}$ and $\vec \nabla \chi$ are orthogonal to $\vec B$,
$\vec B$ must take the form:
\beq
\vec B = A \vec \nabla \chi \times \vec \nabla \eta^{*}.
\enq
However, using \ern{Bcon} we have:
\beq
\vec \nabla \cdot \vec B = \vec \nabla A \cdot (\vec \nabla \chi \times \vec \nabla \eta^{*})=0.
\enq
Which implies that $A$ is a function of $\chi,\eta^{*}$. Now we can define a new comoving function
$\eta$ such that:
\beq
\eta = \int_0^{\eta^{*}}A(\chi,\eta^{'*})d\eta^{'*}, \qquad \frac{d \eta}{d t} = 0.
\enq
In terms of this function we obtain the Sakurai (Euler potentials) presentation:
\beq
\vec B = \vec \nabla \chi \times \vec \nabla \eta.
\label{Bsakurai}
\enq
And the density is now given by the Jacobian:
\beq
\rho = \vec \nabla \mu \cdot (\vec \nabla \chi \times \vec \nabla \eta)
=\frac{\partial(\chi,\eta,\mu)}{\partial(x,y,z)}.
\label{metagecon2}
\enq
It can easily be shown using the fact that the labels are comoving that the above forms of
$\vec B$ and $\rho$ satisfy \ern{Beq}, \ern{Bcon} and \ern{masscon} automatically.

Notice however, that $\eta$ is defined in a non unique way since one can redefine
$\eta$ for example by performing the following transformation: $\eta \rightarrow \eta + f(\chi)$
in which $f(\chi)$ is an arbitrary function.
The comoving coordinates $\chi,\eta$ serve as labels of the magnetic field lines.
Moreover the magnetic flux can be calculated as:
\beq
\Phi = \int \vec B \cdot d \vec S = \int d \chi d \eta.
\label{phichieta}
\enq
In the case that the surface integral is performed inside a load contour we
obtain:
\beq
\Phi (\lambda) = \int_{\lambda} d \chi d \eta= \chi \int_{\lambda} d \eta =\left\{
\begin{array}{c}
 \chi [\eta] \\
  \chi (\eta_{max}-\eta_{min}) \\
\end{array}
\right.
\enq
 There are two cases involved; in one case the load surfaces are topological cylinders;
 in this case $\eta$ is not single valued and hence we obtain the upper value for $\Phi (\lambda)$.
 In a second case the load surfaces are topological spheres; in this case $\eta$ is single valued
 and has minimal $\eta_{min}$ and maximal $\eta_{max}$  values. Hence the lower value of $\Phi (\lambda)$ is obtained.
 For example in some cases $\eta$ is identical to twice the latitude angle $\theta$.
 In those cases $\eta_{min}=0$ (value at the "north pole") and $\eta_{max}= 2 \pi$
 (value at the "south pole").

Comparing the above \eq \ with \ern{chidef} we derive that $\eta$ can be either
{\bf single valued} or {\bf not single valued} and that
its discontinuity across its cut in the non single valued case is $[\eta] =2 \pi$.

The triplet $\chi,\eta,\mu$ will suffice to label any fluid element in three dimensions. But for a non-barotropic
flow there is also another label $s$ which is comoving according to \ern{Ent}. The question then arises of the relation of this
label to the previous three. As one needs to make a choice regarding the preferred set of labels it seems that the physical
ones are $\chi,\eta,s$ in which we use the surfaces on which the magnetic fields lie and the entropy, each label has an obvious
physical interpretation. In this case we must look at $\mu$ as a function of $\chi,\eta,s$. If the magnetic field lines lie on entropy
surface then $\mu$ regains its status as an independent label. The density can now be written as:
\beq
\rho = \frac{\partial \mu}{ \partial s} \frac{\partial(\chi,\eta,s)}{\partial(x,y,z)}.
\label{metagecon3}
\enq
Now as $\mu$ can be defined for each magnetic field line separately according to \ern{metage} it is obvious that
such a choice exist in which $\mu$ is a function of $s$ only. One may also think of the entropy $s$ as a functions
$\chi,\eta,\mu$. However, if one change $\mu$ in this case this generally entails a change in $s$ and the symmetry described in
\ern{metage} is lost.

\section{Lagrangian variational principle of MHD}

Consider the action \cite{metra}:
\ber
A & \equiv & \int {\cal L} d^3 x dt,
\nonumber \\
{\cal L} & \equiv & \rho (\frac{1}{2} \vec v^2 - \varepsilon (\rho,s)) - \frac{\vec B^2}{8 \pi} - \rho \sigma \frac{ds}{dt},
\label{Lagaction2}
\enr
A variation with respect to the Lagrange multiplier $\sigma$ will obviously result in \ern{Ent}. A variation with respect to $s$
will result in:
\ber
\delta_{s} A  & = &  \int d^3 x dt \delta s
[\frac{\partial{(\rho \sigma)}}{\partial t} +  \vec \nabla \cdot (\rho \sigma \vec v)- \rho T]
 \nonumber \\
&+& \int dt \oint d \vec S \cdot \rho \sigma \vec v  \delta s
 -  \int d^3 x \rho \sigma \delta s |^{t_1}_{t_0},
\label{delActions}
\enr
Taking into account the continuity \ern{masscon} we obtain for locations in which the density $\rho$ is not null the result:
\beq
\frac{d \sigma}{dt} =T,
\label{sigmaeq}
\enq
provided that $\delta_{s} A$ vanished for an arbitrary $\delta s$.
Now let us turn our attention to the variation with respect to the fluid element displacement which takes the form:
\ber
\delta A_{\vec \xi} & = & \int \delta {\cal L}_{\vec \xi} d^3 x dt,
\nonumber \\
\delta {\cal L}_{\vec \xi} & = &  \delta \rho (\frac{1}{2} \vec v^2 - w (\rho,s)) - \rho \sigma \delta  \vec v \cdot \vec \nabla s
+ \rho \vec v \cdot \delta  \vec v - \frac{\vec B \cdot \delta \vec B}{4 \pi},
\label{delLagaction2g}
\enr
As most of the terms were calculated previously we will only calculate the term $ - \rho \sigma \delta  \vec v \cdot \vec \nabla s $
which is equal to:
\beq
- \rho \sigma \delta  \vec v \cdot \vec \nabla s = \vec \xi \cdot \rho T \vec \nabla s - \frac{\partial (\rho \sigma \vec \nabla s \cdot \vec \xi)}{\partial t} - \vec \nabla \cdot \left(\rho \sigma (\vec \nabla s \cdot \vec \xi) \vec v\right).
\enq
The above result was obtained using \ern{Ent} and \ern{sigmaeq}. Hence the variation of the action with
respect to a displacement of the fluid elements is:
\ber
\delta_{\vec \xi} A & = & \int d^3 x \rho (\vec v - \sigma \vec \nabla s)  \cdot \vec \xi|^{t_1}_{t_0}
\nonumber \\
&\hspace{-1.7cm} + & \hspace{-1cm}
\int dt \{ \oint d \vec S \cdot [-\rho \vec \xi (\frac{1}{2} \vec v^2 - w (\rho,s))
+ \rho \vec v \left((\vec v - \sigma \vec \nabla s) \cdot \vec \xi\right) + \frac{1}{4 \pi} \vec B \times (\vec \xi \times \vec B)]
\nonumber \\
& \hspace{-1.7cm} + & \hspace{-1cm} \int d^3 x  \vec \xi \cdot [-\rho \vec \nabla w + \rho T \vec \nabla s  - \frac{\partial (\rho \vec v)}{\partial t }
- \frac{\partial (\rho \vec v v_k)}{\partial x_k} - \frac{1}{4 \pi} \vec B \times  (\vec \nabla \times \vec B)
 ]\},
\label{delLagaction2b}
\enr
in which a summation convention is assumed. Taking into account the continuity \ern{masscon} and thermodynamic identities we obtain:
\ber
\delta_{\vec \xi} A & = & \int d^3 x \rho (\vec v - \sigma \vec \nabla s)  \cdot \vec \xi|^{t_1}_{t_0}
\nonumber \\
&\hspace{-1.7cm} + & \hspace{-1cm}
\int dt \{ \oint d \vec S \cdot [-\rho \vec \xi (\frac{1}{2} \vec v^2 - w (\rho,s))
+ \rho \vec v ((\vec v - \sigma \vec \nabla s) \cdot \vec \xi) + \frac{1}{4 \pi} \vec B \times (\vec \xi \times \vec B)]
\nonumber \\
&\hspace{-1.7cm} + & \hspace{-1cm} \int d^3 x  \vec \xi \cdot [- \vec \nabla P - \rho \frac{\partial \vec v}{\partial t}
- \rho (\vec v \cdot \vec \nabla) \vec v  - \frac{1}{4 \pi} \vec B \times  (\vec \nabla \times \vec B ) ]\},
\label{delLagaction3b}
\enr
Hence we obtain the correct dynamical equations for an arbitrary $\vec \xi$. Now suppose that
the equations and boundary conditions hold. Then:
\beq
\delta_{\vec \xi} A =\int d^3 x \rho (\vec v - \sigma \vec \nabla s)  \cdot \vec \xi|^{t_1}_{t_0}
\label{delaction3b}
\enq
If in addition $\vec \xi$ is a small symmetry displacement such that $\delta_{\vec \xi} A =0$ we obtained a conserved Noether current:
\beq
\delta J =\int d^3 x \rho (\vec v - \sigma \vec \nabla s)  \cdot \vec \xi
\label{Noether}
\enq

\section {The labelling symmetry group and its subgroups}

It is obvious that the choice of fluid labels is quite arbitrary. However, when enforcing the $\chi, \eta, \mu$ coordinate
system such that:
\beq
\rho = \frac{\partial(\chi,\eta,\mu)}{\partial(x,y,z)}.
\label{rhojac}
\enq
The choice is restricted to $\tilde \chi, \tilde \eta, \tilde \mu$:
\beq
 {\partial (\tilde \chi, \tilde \eta, \tilde \mu) \over
 \partial (\chi, \eta, \mu)} = 1.
 \label{alphasym}
  \enq
Moreover the Euler potential magnetic field representation:
\beq
\vec B = \vec \nabla \chi \times \vec \nabla \eta,
\label{Bsakurai2}
\enq
reduces the choice further to:
\beq
 {\partial (\tilde \chi, \tilde \eta) \over  \partial (\chi, \eta)} = 1.
 \label{alphasym2}
\enq

\subsection{Metage translations}

In what follows we consider the transformation (see also \ern{metage}):
\beq
\tilde \chi = \chi, \tilde \eta = \eta, \tilde \mu = \mu + a (\chi,\eta)
\enq
Hence $a$ is a label displacement which may be different for each magnetic field line, as the field line
is closed one need not worry about edge difficulties. This transformation satisfies trivially the conditions
(\ref{alphasym},\ref{alphasym2}).
If $a=\delta \mu$ is small we can calculate the associated
fluid element displacement with this relabelling.
\beq
 \vec \xi = -{\partial \vec r \over \partial \mu}\delta \mu = -\delta \mu \frac{\vec B}{\rho}.
\label{ximu}
 \enq
Inserting this expression into the boundary term in \ern{delLagaction3b} will result in:
\beq
\delta A_B =
\int dt  \oint d \vec S \cdot  [\vec B (\frac{1}{2} \vec v^2 - w (\rho,s)) -  \vec v ((\vec v - \sigma \vec \nabla s)\cdot \vec B) ] \delta \mu = 0,
\label{dAB}
\enq
which is the condition for magnetic cross helicity conservation. Inserting \ern{ximu} into \ern{Noether} we obtain the conservation law:
\beq
\delta J =\int d^3 x \rho (\vec v - \sigma \vec \nabla s)  \cdot \vec \xi = -\int d^3 x \delta \mu (\vec v - \sigma \vec \nabla s)  \cdot \vec B
\label{Noether2}
\enq
In the simplest case we may take $\delta \mu$ to be a small constant, hence:
\beq
\delta J = - \delta \mu \int d^3 x (\vec v - \sigma \vec \nabla s) \cdot \vec B = - \delta \mu H_{CNB}
\label{Noether3}
\enq
Where $H_{CNB}$ is the non barotropic global cross helicity \cite{Webb1,Yahalomhel2,Yahalomhel3} defined as:
\beq
H_{CNB} \equiv \int d^3 x (\vec v - \sigma \vec \nabla s) \cdot \vec B  = \int d^3 x \vec v_t \cdot \vec B
\label{Noether4}
\enq
in which $\vec v_t \equiv \vec v - \sigma \vec \nabla s$ is the topological velocity field. We thus obtain
the conservation of non-barotropic cross helicity using the Noether theorem and the symmetry group of metage translations.
Of course one can perform a different translation on each magnetic field line, in this case one obtains:
\beq
\delta J =  -\int d^3 x \delta \mu \vec v_t  \cdot \vec B =
 -\int d \chi d \eta \delta \mu  \oint_{\chi,\eta} d\mu \rho^{-1}  \vec v_t  \cdot \vec B
\label{Noether5}
\enq
Now since $\delta \mu$ is an arbitrary (small) function of $\chi,\eta$ it follows that:
\beq
I = \oint_{\chi,\eta} d\mu \rho^{-1}  \vec v_t  \cdot \vec B
\label{Noether6}
\enq
is a conserved quantity for each magnetic field line. Along a magnetic field line the following equations hold:
\beq
d\mu = \vec \nabla \mu \cdot d \vec r = \vec \nabla \mu \cdot \hat{B} dr = \frac{\rho}{B} dr
\label{Noether7}
\enq
in the above $\hat{B}$ is an unit vector in the magnetic field direction an \ern{metageeq} is used.
Inserting \ern{Noether7} into \ern{Noether6} we obtain:
\beq
I = \oint_{\chi,\eta} dr  \vec v_t  \cdot \hat B = \oint_{\chi,\eta} d \vec r \cdot  \vec v_t.
\label{Noether8}
\enq
which is just the circulation of the topological velocity along the magnetic field lines. This quantity can be
written in terms of the generalized Clebsch representation of the velocity \cite{Yahalom1}:
\beq
\vec v =  \vec \nabla \nu + \alpha \vec \nabla \chi + \beta \vec \nabla \eta + \sigma \vec \nabla s, \qquad \vec v_t =  \vec \nabla \nu + \alpha \vec \nabla \chi + \beta \vec \nabla \eta
\label{vform}
\enq
as:
\beq
I  = \oint_{\chi,\eta} d \vec r \cdot  \vec v_t = \oint_{\chi,\eta} d \vec r \cdot  \vec \nabla \nu = [\nu].
\label{Noether9}
\enq
$[\nu]$ is the discontinuity of $\nu$. This was shown to be equal to the amount of non barotropic cross helicity per unit
 of magnetic flux \cite{Yahalomhel2,Yahalomhel3}.
\begin{equation} \label{loc_}
I=[\nu]= \frac{dH_{CNB}}{d \Phi}.
\end{equation}

\subsection{Transformations of magnetic surfaces}

Consider the following transformations:
\beq
 \tilde{\eta} = \eta + \delta \eta(\chi,\eta), \qquad \tilde{\chi} =  \chi + \delta \chi(\chi,\eta), \qquad \tilde{\mu} = \mu
\enq
in which $\delta \eta,\delta \chi$ are considered small in some sense. Inserting the above quantities into
\ern{alphasym2} and keeping only first order terms we arrive at:
\beq
\partial_{\eta} \delta \eta + \partial_{\chi} \delta \chi = 0.
\label{delchieteq}
\enq
This equation can be solved as follows:
\beq
 \delta \eta = \partial_{\chi} \delta f, \qquad  \delta \chi = -\partial_{\eta} \delta f,
 \label{delf}
\enq
in which $\delta f= \delta f (\chi,\eta)$ is an arbitrary small function. In this case we obtain a particle displacements of the
form:
\beq
 \vec \xi = -{\partial \vec r \over \partial \chi}\delta \chi -{\partial \vec r \over \partial \eta}\delta \eta =
 -\frac{1}{\rho} \left( \vec \nabla \eta \times \vec \nabla \mu \ \delta \chi +  \vec \nabla \mu \times \vec \nabla \chi \ \delta \eta  \right)
 = \frac{\vec \nabla \mu}{\rho} \times ( \vec \nabla \eta \delta \chi - \vec \nabla \chi \delta \eta )
\label{xi2}
 \enq
A special case that satisfies \ern{delchieteq} is the case of a constant $\delta \chi$ and  $\delta \eta$, those two independent displacements
lead to two new topological conservation laws:
\beq
\delta J_\chi =  \delta \chi \int d^3 x \vec v _t \cdot \vec \nabla \mu \times \vec \nabla \eta= \delta \chi\  H_{CNB \chi},
\quad
\delta J_\eta =  \delta \eta \int d^3 x \vec v _t \cdot \vec \nabla \chi \times \vec \nabla \mu = \delta \eta\  H_{CNB \eta}.
\label{Noether10}
\enq
Where the new non barotropic global cross helicities are defined as:
\beq
H_{CNB \chi} \equiv \int d^3 x \vec v _t \cdot \vec \nabla \mu \times \vec \nabla \eta, \quad
H_{CNB \eta} \equiv \int d^3 x \vec v _t \cdot \vec \nabla \chi \times \vec \nabla \mu
\label{Noether11}
\enq
We remark that those topological constants will only be conserved under special boundary conditions satisfying:
\ber
 \oint d \vec S \cdot \left[-\rho \vec \xi (\frac{1}{2} \vec v^2 - w (\rho,s))
+ \rho \vec v (\vec v _t \cdot \vec \xi) + \frac{1}{4 \pi} \vec B \times (\vec \xi \times \vec B)\right] = 0
\label{boundary}
\enr
for:
\beq
 \vec \xi_\chi =   \frac{1}{\rho} \left( \vec \nabla  \mu \times \vec \nabla  \eta \ \right) \delta \chi, \qquad
 \vec \xi_\eta =   \frac{1}{\rho} \left( \vec \nabla  \chi \times \vec \nabla  \mu\ \right) \delta \eta
\label{xi3}
 \enq
This is more plausible for magnetic field lines which lie on topological torii. In this case $\eta$ is non single valued \cite{YaLy} and thus
the translation in this direction resembles moving fluid elements along closed loops. Finally we remark that for barotropic MHD $\vec v_t$
can be replaced with $\vec v$.

\section{Conclusion}

We have shown the connection of the translation symmetry groups of labels to both
the global non barotropic cross helicity conservation law and the conservation law of circulations of topological velocity along
magnetic field lines. The latter were shown to be equivalent to the amount of non barotropic cross helicity per unit
 of magnetic flux \cite{Yahalomhel2,Yahalomhel3}. Further more we have shown that two additional cross helicity conservation laws exist the $\chi$ and $\eta$ cross helicities. Possible applications for MHD  constraints of the current constants of motion are described in \cite{Yahalomhel3}. The importance of constants of motion for stability analysis is also discussed in \cite{Katz}.

\end{document}